# Graphene Nucleation on Transition Metal Surface: Structure Transformation and Role of the Metal Step Edge

*Junfeng Gao,[1,2] Joanne Yip,[1] Jijun Zhao,[2] Boris I. Yakobson,[1,3,4] Feng Ding[1,3*]*

1) Institute of Textiles and Clothing, Hong Kong Polytechnic University, Hong Kong, China

2) Key Laboratory of Materials Modification by Laser, Ion and Electron Beams (Dalian University of Technology), Ministry of Education, Dalian 116024, China

3) ME&MS Department, Rice University, Houston, TX 77005, USA

4) Department of Chemistry, Rice University, Houston, TX 77005, USA

E-mail: feng.ding@rice.edu

**Abstract**. The nucleation of graphene on a transition metal (TM) surface, either on a terrace or near a step edge, is systematically explored using density functional theory (DFT) calculations and applying the two-dimensional (2D) crystal nucleation theory. Careful optimization of the supported carbon clusters, $C_N$ (with size N ranging from 1 to 24), on the Ni(111) surface indicates a ground state structure transformation from a one-dimensional (1D) C chain to a two-dimensional (2D) $sp^2$ C network at N ~ 10-12. Furthermore, the crucial parameters controlling graphene growth on the metal surface, nucleation barrier, nucleus size, and the nucleation rate on a terrace or near a step edge, are calculated. In agreement with numerous experimental observations, our analysis shows that graphene nucleation near a metal step edge is superior to that on a terrace. Based on our analysis, we propose the use of seeded graphene to synthesize high-quality graphene in large area.

1. Introduction

Graphene has drawn the attention of physicists, chemists, and material scientists in a short period since its experimental synthesis in 2004.[1] This thinnest mono-atomic fabric is extremely strong, with a Young's modulus of > 1000 GPa and a strength of > 100 GPa.[2] Its measured thermal conductivity is close to or even in excess of that of diamond.[3] A tunable band gap emerges after cutting the two-dimensional (2D) one-atom-thick semi-metal into graphene nanoribbons (GNR)[4] via chemical functionalization[5] or physisorption of molecules.[6] Due to its exceptional electronic, mechanical, thermal, and optical performance, numerous potential applications have been proposed.[7] For example, researchers have proposed that graphene be used in future as a replacement material for silicon (Si) once

Si-based electronic technology has reached its quantum limit.[8] Graphene has further been widely investigated as a key material in multi-functional composites,[9] sensors,[10] flexible and transparent electrodes,[11] super-capacitors,[12] and for other purposes.[4c,13] For all of these potential applications, the synthesis of high-quality thin graphene layers on a large scale is highly desired.[14]

Motivated by these numerous potential applications and the great demand for high-quality graphene, many synthesizing methods have been developed and explored extensively in the past few years. The most used methods include (i) Scotch Tape mechanical peeling,[1a,1c] (ii) the sublimation of SiC at high temperatures,[15] (iii) the intercalation of graphite,[16] (iv) chemically functionalized graphene reduction,[17] and (v) transition metal (TM)-catalyzed chemical vapor deposition (CVD).[18] Of these, CVD graphene synthesis stands out for its numerous advantages: (i) it can be achieved at a relatively low temperature (i.e., 1000 K or lower, which is notably lower than the temperature required for SiC sublimation, i.e., 1500-2000 K), (ii) single- or few- layer graphene of very high quality can be synthesized easily due to the catalyst-assisted defect healing, (iii) synthesis of graphene of a very large area is possible (e.g., 100-1000 square inches), and (iv) synthesized graphene can be easily transferred into other substrates for further processing.

To improve the quality and scale of synthesized graphene, extensive efforts have been devoted to gaining a deep understanding of the mechanisms of graphene nucleation and growth in CVD experiments.[18a,b,18d-m,19] The TM-graphene interaction has been found as a crucial parameter for growth control. More specifically, theoretical calculations and experimental characterizations have revealed a very strong TM-C interaction on the graphene edge[20] and thus the formation of dome-like graphene islands on the metal surface.[19a] A systematic experimental study carried out by McCarty et al. revealed that graphene nucleation tends to occur near a metal step edge only at low C concentration while simultaneous nucleation both near step edge and on a terrace occurs at a very high C concentration.[19b] This implies a higher nucleation barrier to initiating a graphene nucleus on metal terrace than that near a metal step edge and the barrier difference can be lowered by increasing the C concentration. Although there is an extensive body of experimental work in this area, theoretical explorations are relatively rare. Chen et al. very recently investigated the formation of a C dimer on a metal surface as the very initial graphene nucleation stage.[20a] Theoretical research carried out by Saadi et al. showed a surprisingly small nucleation barrier (i.e., G* < 2 eV) and nucleus size (the energy maximum appearing at N* = 1-6).[20b] Amara et al. revealed the TM-assisted defect healing of graphene in their CVD synthesis.[21]

It is worth to note that, although numerous TMs have been used in graphene synthesis, their roles in graphene CVD growth are very similar. Recently both theoretical studies[20a,22] and experimental researches[18h,19b,23] have clearly shown the similarities among these TMs. So, we believe the study of graphene growth on one of the most used metal Ni also sheds light on the using of other TMs in graphene synthesis.

Here we report a systematic study of graphene nucleation on a Ni(111) surface. Our results demonstrate that C chain formation on a TM surface is very stable at small and intermediate sizes and that a ground state structure transformation from the C chain to the $sp^2$ C network occurs at N = 12 and 10 for the $C_N$ clusters on the Ni(111) terrace and near the step edge, respectively. The most stable $sp^2$ C networks always have a few (i.e., one to three) pentagons in their formation, which originates from the requirement to reduce the number of edge C atoms. Further analysis showed that the transformation from the C chain to the C $sp^2$ network play a crucial role in the nucleation of graphene. This transformation also results in a linear reduction of the nucleation barrier and a constant nucleus size in a broad range of Δμ, which is the C chemical potential deduction from feedstock to graphene. Compared to that on a Ni(111) terrace, the nucleation of graphene near a step edge has a significantly lower nucleation barrier (i.e., ~ 1.0-2.0 eV lower), and the difference between these two situations increases dramatically with a smaller Δμ. Thus, nucleation near the metal step edge dominates the graphene nucleation process at a small Δμ, and spontaneous graphene nucleation on both the terrace and near the step edge will occur at a large Δμ.

## 2. Results and discussion

### 2.1. Graphene edge formation energy

Let us first address the difference between graphene nucleation on a metal terrace and near a metal step edge. As shown in Figure 1, both free zigzag (ZZ) and armchair (AM) graphene edges have very large formation energies: 13.46 and 10.09 eV/nm, respectively (Figure 1a, 1d). When a graphene edge approaches a Ni(111) surface, its formation energy is notably reduced because of the strong binding between the edge C atoms and the active metal surface (Figure 1b, 1e). The formation energy of a ZZ edge is reduced nearly 50% to 6.95 eV/nm (Figure 1b) and that of the AM edge drops by about 30% to reach a similar value, 7.14 eV/nm (Figure 1e). The AM edge's relatively lower formation energy reduction can be explained by its stable triple bonds. A further reduction in formation energy occurs when a graphene edge approaches a metal step edge on the Ni(111) surface (Figure 1c, 1f). The formation energies of ZZ and AM graphene edges on a metal step edge, for example, are only 3.45 and 4.79 eV, respectively. This significant reduction in edge formation energy implies that graphene nucleation can be facilitated by a metal step edge, which is in agreement with a wide range of experimental observations.[18l,19b,24]

As illustrated in Figure 1, the formation energies of graphene edges on a Ni(111) terrace and near a step edge can be used to model the formation energy of a graphene patch and to calculate the nucleation barrier and nucleus size as a function of the chemical potential drop, Δμ. However, the behavior of graphene patches of a very small size, namely, carbon clusters, may be very different from that of regular patches (i.e., a pure hexagonal sp$^2$ C network with well-defined ZZ or AM edges). For example, a one-dimensional (1D) C chain or ring is much more stable than a sp$^2$ C network at a small size (n < 20)[25] in vacuum, and C chain formation with both ends tightly attached to the metal surface has been widely observed in previous molecular dynamic and Monte Carlo simulations.[21,26] To address the unique formation of small C clusters, we performed a global search of the ground states of supported carbon clusters as the first step in modeling the formation energy of the C clusters on a metal surface.

### 2.2. Formation of $C_N$ (N=1, .., 24) clusters on Ni(111) terrace and near a metal step edge

Carbon clusters $C_N$ (N = 1, … , 24) supported on a Ni(111) surface (noted as $C_N$@Ni(111) hereafter) were optimized using the conjugate gradient method, which is implemented in the Vienna *ab initio* Software Package (VASP)[27] (detailed information on the method of calculation provided in the Supporting Information [**SI**]). For each cluster of N > 10, more than ten different configurations were explored, and the most stable one was taken as the ground state. **Figure S1** in the **SI** shows all structures of $C_{14}$@Ni(111) that were explored and their corresponding formation energies. These optimized structures can be classified into three categories: (i) C ring (Figure2 c. **C$_9$-3**), (ii) C chain (Figure2 a. **C$_9$-G**), and (iii) sp$^2$ C networks that are primarily formed by 5-, 6-, or 7-membered rings (Figure 2). On the Ni(111) terrace, the energy order of these supported clusters is very different from that of their free-standing counterparts. For the free C clusters, ring formation dominates the ground states in the intermediate-sized (N ~ 6-20) clusters, and closed sp$^2$ networks with 12 pentagons or fullerenes dominate those in the larger clusters (N > 20).[25] On a Ni(111) terrace, a C ring is always less stable than the corresponding C chain (e.g., a. **C$_9$-G** versus c. **C$_9$-3** in Figure 2) because the passivation of the two free ends of a C chain on a metal surface significantly reduces its formation energy. Our calculations show that the formation energy of the end of a C chain is reduced from ~ 3.5 eV/end to ~ 0.20 eV/end upon Ni(111) terrace passivation.

It is very surprising that the ground state structures of all of the $C_N$ networks explored in this study (10 < N < 24) have one to three pentagons and that the energies of the pure hexagonal networks are significantly larger. The energy differences are from 1.3 to 2.4 eV, as shown in Figure 2 (**e.g.,** e. **C$_{10}$-H** **versus** d. **C$_{10}$-1;** g. **C$_{13}$-H versus** f. **C$_{13}$-G; and** o. **C$_{24}$-H versus** n. **C$_{24}$-G**). This finding is in striking contrast to a previous interpretation of graphene dome formation,[19a] and can be more or less explained as follows. The formation energy of a sp$^2$ C network comes primarily from its edge atoms, and thus a reduction in the number of these atoms is energetically preferred. Incorporating one or few pentagons into a sp$^2$ C network alters its shape from flat to bowl-like, which normally results in a reduced circumference length or number of edge atoms.

Figure 3 shows some of the most stable C$_N$ cluster structures near a step edge along the (110) direction on a Ni(111) surface. Due to the enhanced activity of the metal step edge, upon optimization, C$_N$ clusters tend to have more C atoms attached to the metal step. The most stable clusters therefore exhibit a partial moon shape, which is in sharp contrast to the circular shape of the C$_N$ clusters on a Ni(111) terrace (e.g., the ground states of j. C$_{20}$-n. C$_{24}$ in Figure 2).

Figure 4a shows the formation energies of the C ring, chain, and most stable sp$^2$ networks on a Ni(111) terrace (partial data shown in Figure 2) and those of the C chain and most stable sp$^2$ networks near the metal step edge (partial data shown in Figure 3) as a function of cluster size, N. The formation energy is defined as

$$E_N = E(C_N@Ni) - E(Ni) - N \times \epsilon_G, \tag{1}$$

where E(C$_N$@Ni(111)) is the energy of a C$_N$ cluster on a Ni substrate, E(Ni) is the energy of the Ni substrate, and $\epsilon_G$ is the energy per carbon atom of graphene.

On the terrace, the formation energy of the 1D C chain increases linearly with cluster size N, and these data can be fitted as

$$E_{ch}(\text{Terrace}) = 0.81 \times N + 0.40 \text{ eV}, \tag{2}$$

where the energy increment of ~ 0.81 eV is roughly the energy difference between a sp$^1$ hybridized C atom and a sp$^2$ hybridized C, and the second term on the right-hand side of the formula, 0.40 eV, is the formation energy of the two chain ends that are passivated by the Ni(111) surface. It can be seen that the formation energy of a chain end is notably reduced from ~ 3.5 eV/end to 0.2 eV/end upon TM passivation. Compared with the ring formation, the chain formation has the significant advantage of less curvature energy and the negligible disadvantage of end formation energy (a ring has no end). In vacuum, the large end formation energy destabilizes chain formation, whereas on a metal surface, the notably reduced such energy lowers the energy of chain formation, and thus a C ring is never the most stable structure on a Ni(111) surface.

The formation energy of a $sp^2$ network is greater than that of a C chain at small sizes (N < 12) because of the large portion of its edge C atoms. However, a $sp^2$ network has the advantage of $sp^2$-hybridized C atoms, whose energy is significantly lower than that of a $sp^1$ C chain. Thus, the $sp^2$ network eventually becomes the most stable configuration beyond the critical size, i.e., $N_C$ = 12. The formation energy of the most stable $sp^2$ network can be fitted as

$$E_{sp2}(\text{Terrace}) = 2.4\ N^{1/2} + 1.6\ \text{eV}. \qquad (3)$$

Similar to the $C_N$ clusters on a terrace, a structural transformation from a C chain to a $sp^2$ C network appears in the energy plot of C clusters near a metal step edge at $N_C$ = 10 (Figure 4a). The formation energies of the C chain near this edge can be fitted as

$$E_{ch}(\text{Step}) = -0.263 + 0.775 \times N\ \text{eV}, \qquad (4)$$

where the chain end formation energy is further reduced to −0.13 eV/end, and the formation energy increment changes slightly, i.e., by 3%, both demonstrating the enhanced chemical activity of the step edge. The formation energy of a $sp^2$ C network near a step edge can be fitted as

$$E_{sp2}(\text{Step}) = 1.992 \times N^{1/2} + 1.328\ \text{eV}. \qquad (5)$$

Figure 4b shows the formation energy difference between the C clusters on a Ni(111) terrace and those near a metal step edge as a function of cluster size. Clearly, approaching a step edge always stabilizes the C cluster. The energy difference rises to 2 eV or higher at a size of N > 12. As we will see later, such an energy difference is crucial in the graphene nucleation behavior displayed on a terrace or near a step edge.

**2.3. Graphene nucleation barriers and nucleation rates on Ni(111) terrace and near a metal step edge**

During crystal nucleation or growth, the change in Gibbs free energy as a function of the number of atoms in the crystalline phase, G(N) = E(N) − Δμ × N,[28] where Δμ is the chemical potential difference between this phase and the atom source, dominates the behavior of both nucleation and growth. The nucleus size and nucleation barrier, (N*, G*), are defined as the maximum of the G(N) curve. Following this definition, we can easily determine G* and N* as a function of Δμ for graphene nucleation on a Ni(111) terrace or near a step edge from Eqs. (2)-(5).

Figure 5a presents nucleation barrier G* and nucleus size N* as a function of Δμ. In both cases, the nucleus size exhibits stepwise behavior that stems from the ground state structure transformation from the C chain to the sp$^2$ network (see **S2** in the **SI** for further details). For nucleation on a terrace (near the step edge), N* = 12 (10) in the range of Δμ ∈ [0.346 eV, 0.81 eV] ([0.315 eV, 0.775 eV]). The nucleation barrier decreases linearly with Δμ in this regime, that is, from 5.77 eV to 0.2 eV for nucleation on a terrace and from 4.47 eV to 0.0 eV for that near a step edge. In the region of Δμ > 0.81 eV for nucleation on a terrace or Δμ > 0.775 eV for that near a step edge, the nucleation barrier goes to zero and the nucleus size drops to N* = 1 abruptly. This absence of nucleation barrier and very small nucleus size imply that graphene nucleation may occur with a deposited carbon cluster of any size. In this case, graphene nucleation or growth is dominated by the C deposition rate and C diffusion on the metal surface, and spontaneous nucleation and growth will occur, although this must be a rare situation in the CVD growth of graphene, when the driving force, Δμ, is so large (~ 0.8 eV). In a low Δμ regime (Δμ < 0.346 and Δμ < 0.315 for nucleation on a terrace and near a step edge, respectively), typical 2D nucleation is displayed, where both the nucleation barrier and nucleus size increase significantly with a decrease in Δμ, G* ~ 1/μ and N* ~ 1/μ$^2$. At Δμ = 0.2 eV, G* reaches 8.8 eV/6.3 eV and N* reaches 36/25 for nucleation on the terrace/near the step edge. This very high nucleation barrier indicates that nucleation will rarely occur in this region.

From classical nucleation theory,[28] the 2D nucleation rate of graphene on Ni(111) surface can be estimated as

$$R_{nul} = R_0 \exp(-G^*/k_b T), \qquad (6)$$

Where $k_b$ is the Boltzmann constant and the pre-factor $R_0$ can be appropriately estimated as $R_0 \sim 4 * 10^{21}$ cm$^{-2}$ s$^{-1}$ (see **S3** in the **SI**). Figure 6a shows the nucleation rate of graphene on a terrace or near a step edge as a function of $\Delta\mu$ at several typical experimental temperatures: 873 K, 1073 K, and 1273 K. It is clear that the nucleation rate is extremely sensitive to both temperature and $\Delta\mu$. A slight variation in temperature or $\Delta\mu$ may result in a dramatic change in the growth rate. For example, at 1073 K, altering the $\Delta\mu$ from 0.4 to 0.6 eV leads to a growth rate change on the terrace, $R_T$, of 12 orders of magnitude (from $10^{-4}$ to $10^8$ cm$^{-2}$s$^{-1}$). Varying the temperature by 200 K results in a six or more order-of-magnitude change in the $R_T$ or $R_E$.

The differences in graphene nucleation on a terrace and near a step edge at these temperatures are shown in Figure 6b. As expected, the $R_E/R_T$ ratio changes monotonically with either temperature or $\Delta\mu$, and the difference vanishes at $\Delta\mu = 0.81$ eV, where any C monomer may initiate graphene nucleation or N* = 1. In a typical growth region, $\Delta\mu \in$ [0.3 eV, 0.65 eV] at 1073K, and $R_E/R_T$ varies from $10^4$ to $10^8$, depending on the temperature. However, the great advantage of $R_E$ is that it does not imply that nucleation must start from the step edge. Because the effective area of a step edge is only one or two atoms in width, the terraces of a crystal may have the huge advantage of a large area; for example, the typical distance between two neighboring step edges is a few to a few tens of microns. Hence, the effective area ratio, $A_T/A_E$, may reach $10^4$ to -$10^{-5}$, and thus the probability of nucleating the first nucleus on area ($A_T/A_E$) * ($R_T/R_E$) is 10 to $10^{-4}$, depending on $\Delta\mu$. Although nucleation near the step edge is preferred in most parameter spaces of temperature, $\Delta\mu$, and the average distance between neighboring step edges, we can see that nucleation on the terrace may be preferable at a large $\Delta\mu$, a high temperature, and large neighboring step edge distance. This conclusion is in good agreement with a recent observation that nucleation on a terrace occurs at a high degree of C monomer coverage on the metal surface, $N_1$.[19b] Because $\Delta\mu \sim \ln (N_1)$, a high degree of C monomer coverage means a large $\Delta\mu$.

A very high nucleation rate will result in the simultaneous formation of many nuclei on the TM surface, and, later on, the coalescence of these independently nucleated and grown graphene islands will result in numerous linear defects or grain boundaries. Thus, a relatively low nucleation rate is preferred for high-quality graphene growth. As shown in Figure 6, a low nucleation rate can be achieved through the use of a lower temperature and a low $\Delta\mu$ (i.e., 0.3-0.5 eV).

As can be seen in Figure 5, the nucleation barrier on the terrace and near the step edge is 8.8 and 6.3 eV, respectively, when $\Delta\mu$ = 0.2 eV. A high nucleation barrier notably diminishes the probability of graphene nucleation. This observation points toward another means of growing single crystal graphenes, seeded growth, which is a well-known trick for large single crystal growth in 3D.[29] Adding a small graphene patch onto the metal surface can help to avoid the nucleation stage of graphene growth, and multi nucleation sites are consequently prohibited by the high nucleation barrier. In normal growth conditions, $\Delta\mu$ = 0.2 eV is two to three times larger than thermal activation energy, $k_bT$, and thus there is still sufficient driving force for the graphene seeds to grow large.

### 3. Conclusion

In conclusion, we have investigated the nucleation of graphene on a Ni(111) terrace and near a step edge using a multi-scale approach. The structural optimization of small graphene patches or carbon clusters (1 ⩽ N ⩽ 24) based on a DFT potential energy surface reveals a notable structure transformation from a C chain to a $sp^2$ C network at N = 12 and 10 for C clusters on the terrace and near the step edge, respectively. Furthermore, incorporating a small number of pentagons (1-3) into a $sp^2$ network is found to reduce the formation energy significantly. We realize that the formation energies of C clusters can in principle be computed more precisely by Quantum Monte Carlo (QMC)[30] than by the GGA. However, QMC is extremely computationally expensive, and the structural optimization of hundreds of metal-C configurations studied here is simply unfeasible. Previous work[30] shows the $C_{20}$ isomer formation energies obtained by GGA and QMC differ by ~1.0 eV, notably less than the formation energy of a $C_{20}$ on Ni(111) surface (~10 eV). Consequently, the GGA method appears sufficient for revealing the general trends in focus of present study. Based on these DFT computations, we further calculate the nucleation barrier, nucleus size, and nucleation rate of graphene on a Ni(111) surface as a function of $\Delta\mu$ based on crystal growth theory. It is found that nucleation near a step edge has a significantly lower barrier (i.e., 2 eV) than that on a terrace and that the nucleation rate near the former may be $10^4$-$10^7$ times greater than that on the latter. Nucleation near the step edge is expected to dominate graphene nucleation in most cases, unless it occurs on a very flat surface and at a high temperature and with a large Δμ. Based on the observation of the very large nucleation barrier, we have proposed a strategy to grow large-area single crystal graphene on TM, that is, seeded growth, with its feasibility proved in a normal growth condition (i.e., T = 872-1272 K and Δμ = 0.2 eV). The deeper insights into the atomistic nucleation mechanisms of CVD graphene growth presented herein are expected to guide the growth of high-quality graphene for numerous applications.

**Acknowledgement.** This work was supported by Hong Kong Polytechnic University funds (1-ZV3B; A-PD1U; A-PH93; A-PJ50) and the Fundamental Research Funds for the Central Universities of China (No. DUT10ZD211). Work at Rice was supported by the Office of Naval Research MURI project.

## TOC Graphics

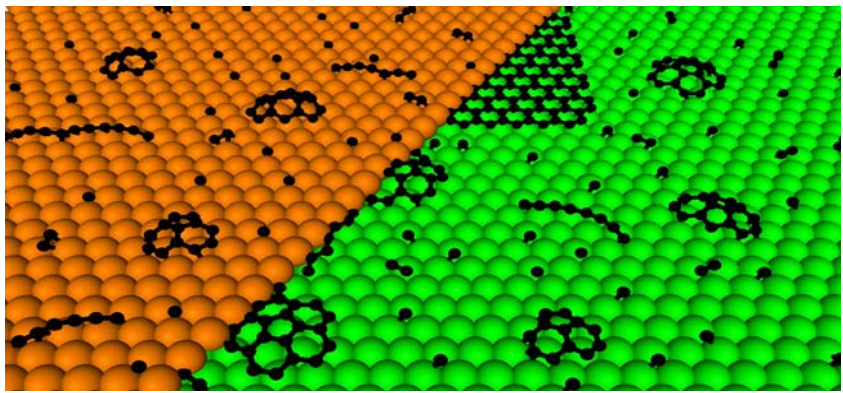

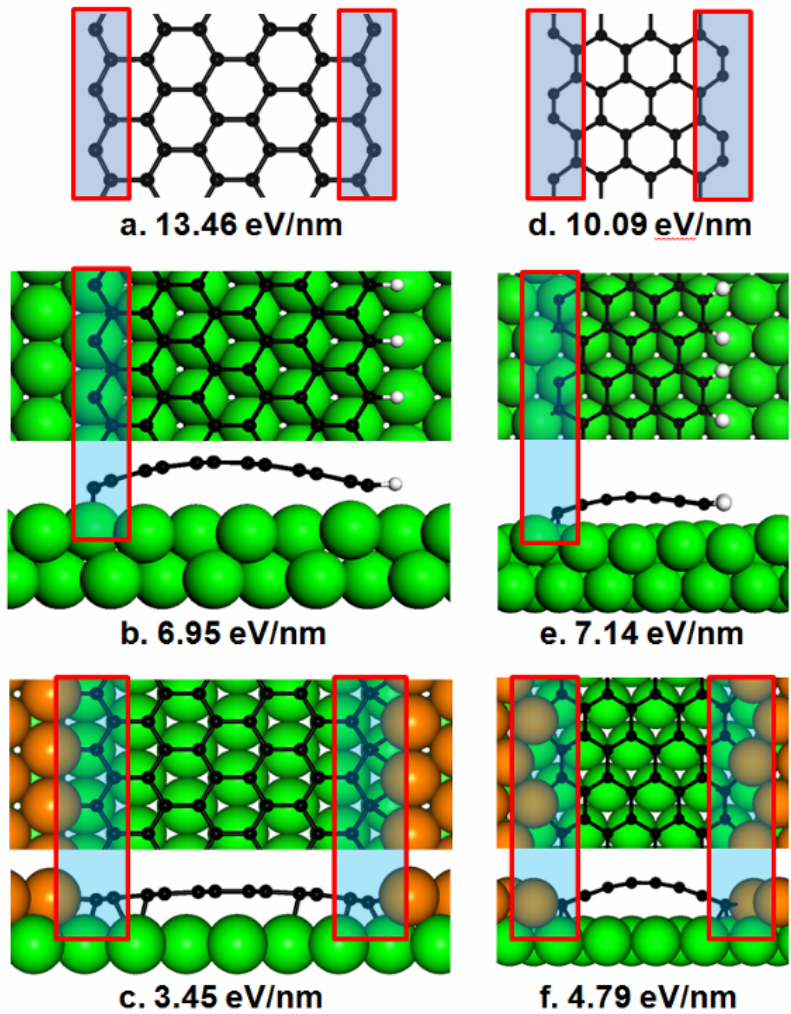

Figure 1. Optimized armchair (AM) and zigzag (ZZ) graphene edges in vacuum (**a, d**), on a Ni (111) terrace (**b, e**), and near a metal step (**c, f**), along with the corresponding formation energies for each structure.

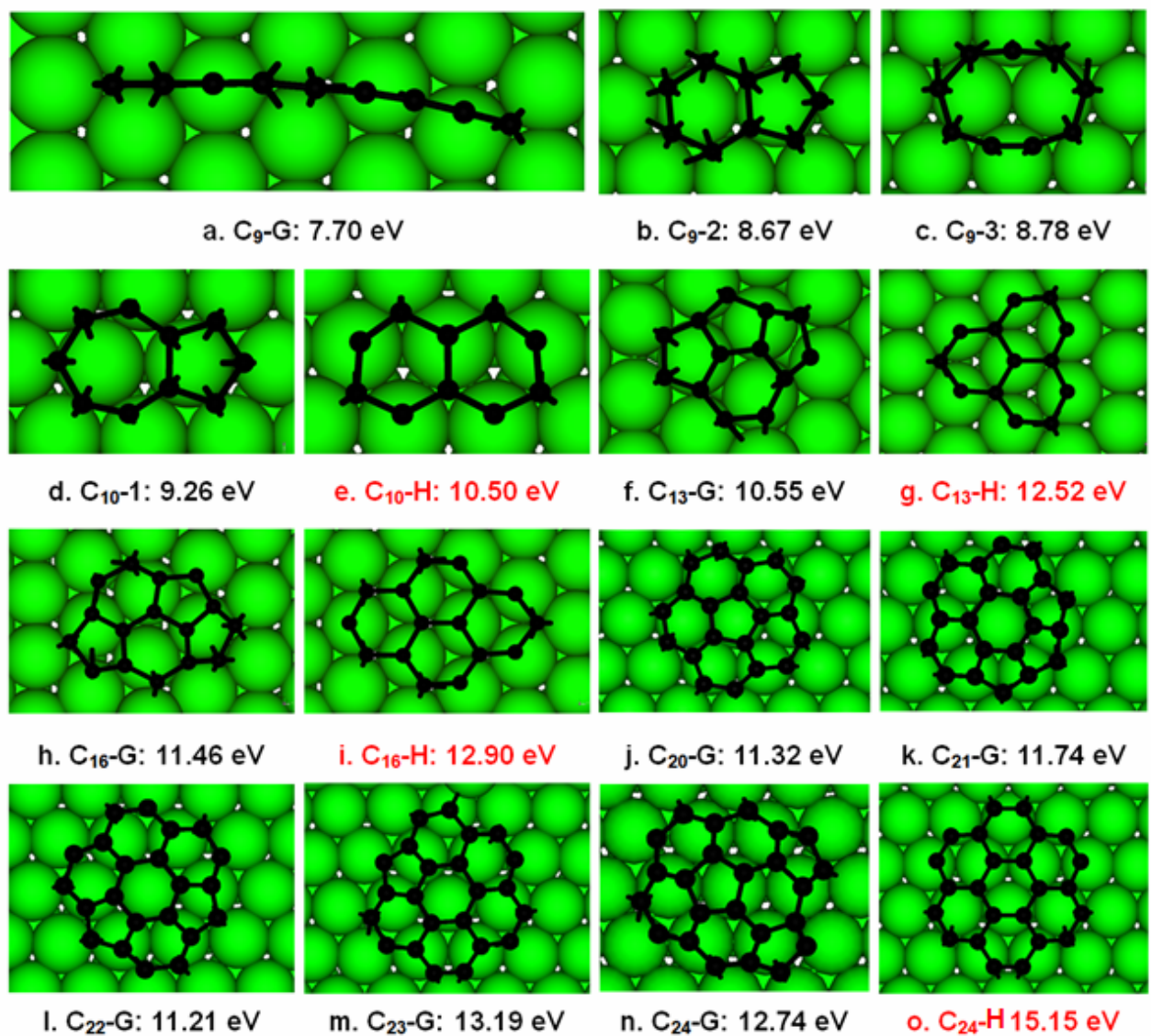

Figure 2. Optimized C clusters supported on a Ni(111) terrace: the graphs marked by "$C_N$-G" are the ground states, and those marked by "$C_N$-H" are hexagon-only structures. Their corresponding formation energies are also given.

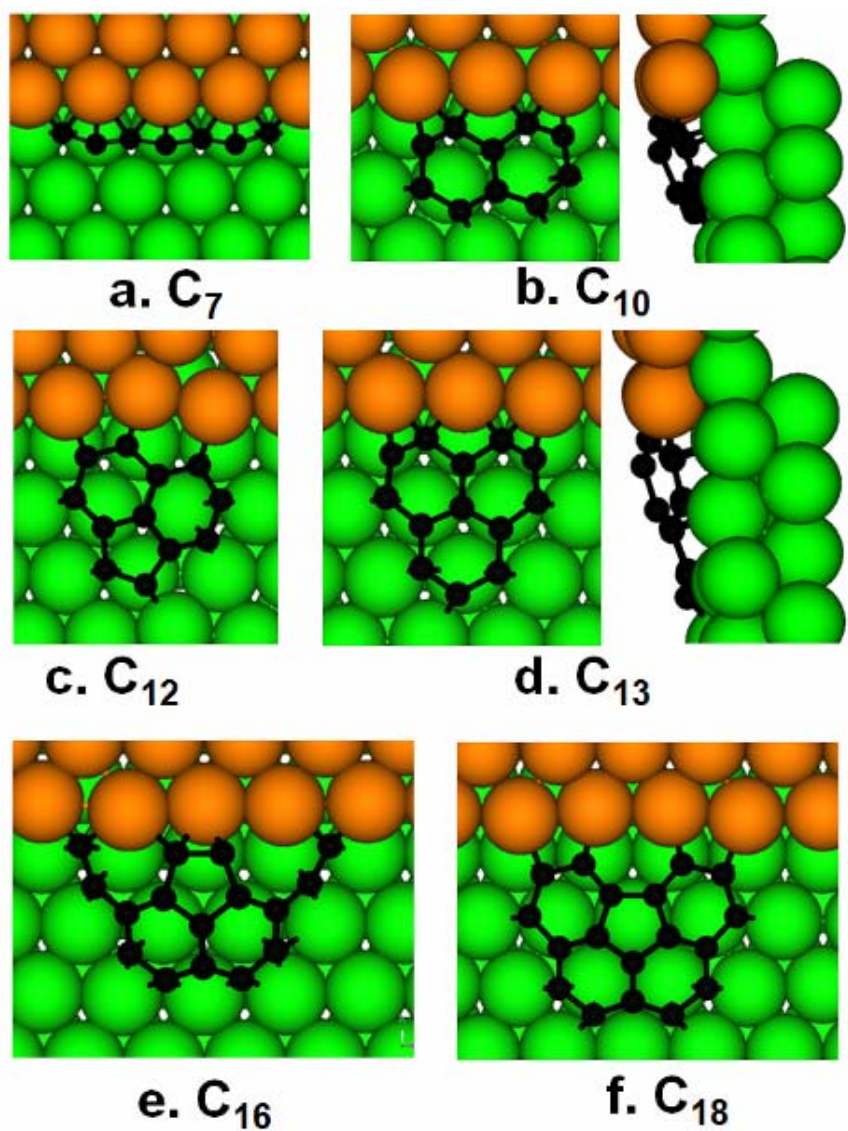

Figure 3. Typical optimized structures of supported $C_N$ clusters near a Ni step edge on the (111) surface. Both the top and side views of the $C_{10}$ and $C_{13}$ clusters are shown.

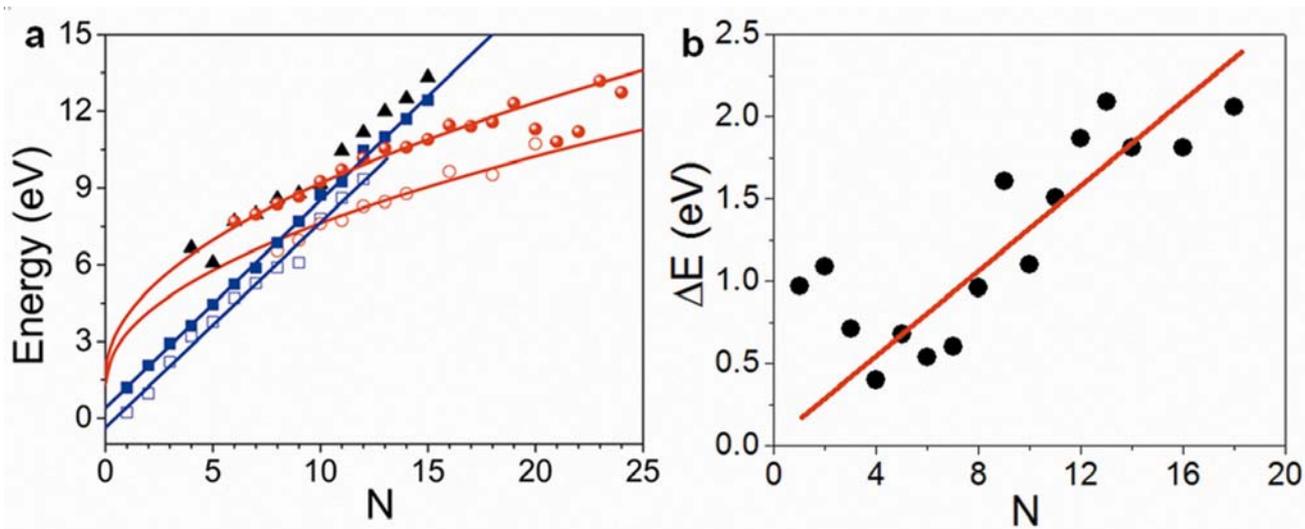

Figure 4. (a) Energy of supported $C_N$ clusters versus cluster size, N, on a Ni(111) terrace and near a step edge. The energies of the C chains, rings, and $sp^2$ networks are shown in the squares, triangles, and circles, respectively. The solid and hollow symbols represent the energies of the C clusters on the terrace and near the metal step, respectively. The straight lines and curves are fitted with Eqs. (2)-(5). (b) The energy difference between the optimized $C_N$ on the Ni(111) terrace and near the step edge, with the straight line providing a linear fit to the data to guide the eye.

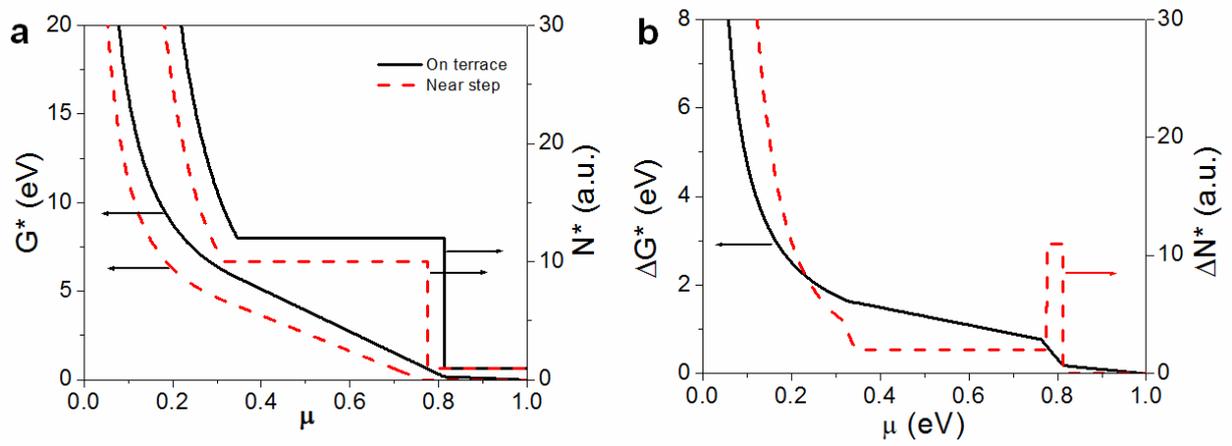

Figure 5. (a) Nucleus size N* and nucleation barrier G* as a function of the graphene nucleation/growth driving force or chemical potential difference between the C in graphene and C in feedstock. (b) The difference between the nucleation barriers and nucleus sizes as a function of the chemical potential: ΔG* versus Δμ and ΔN* versus Δμ.

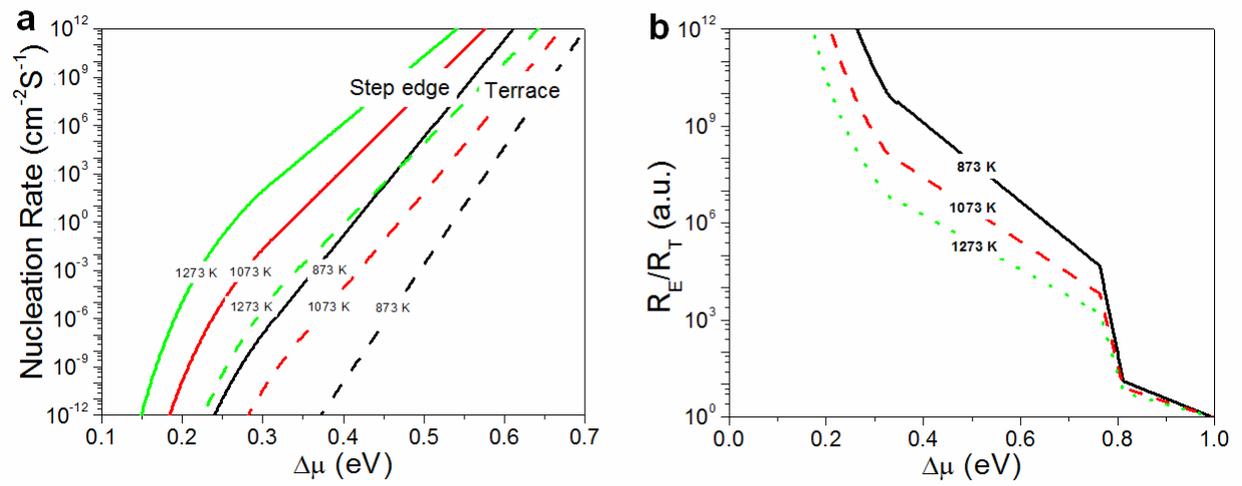

Figure 6. (a) Nucleation rates of graphene growth on a Ni(111) terrace, $R_T$, or near a step edge, $R_E$, as a function of $\Delta\mu$ at temperatures of 873 K, 1073 K, and 1273 K; (b) their ratio, $R_E/R_T$.